\documentclass[prd,twocolumn,superscriptaddress,floatfix]{revtex4}
\usepackage[utf8]{inputenc}
\usepackage{graphicx}
\usepackage{graphics,bm}
\usepackage{amssymb}
\usepackage{color}
\usepackage{palatino}

\newcommand{\bra}[1]{\langle #1 |}
\newcommand{\ket}[1]{| #1 \rangle}
\newcommand{\braket}[2]{\langle #1 | #2 \rangle}
\def\6{\langle}
\def\9{\rangle}

\newcommand\al{\alpha}

\newcommand\wtPsi{\widetilde{\mathtt{\Psi}}}
\newcommand\wtPi{\widetilde{\mathtt{\Pi}}}

\newcommand\wta{\widetilde{\mathtt{a}}}

\newcommand{\be}{\begin{equation}}
\newcommand{\ee}{\end{equation}}
\newcommand{\ba}{\begin{eqnarray}}
\newcommand{\ea}{\end{eqnarray}}

\usepackage{amsmath}

\begin{document}
%opening
\title{Canonical field quantization in momentum space and quantum space-time}
\author{Lucas C. C\'{e}leri}
\email{lucas@chibebe.org}
\affiliation{Instituto de F\'{i}sica, Universidade Federal de Goi\'{a}s, Caixa Postal 131, 74001-970, Goi\^{a}nia, Brazil}

\author{Vasileios I. Kiosses}
\email{kiosses.vas@gmail.com}
\affiliation{Instituto de F\'{i}sica, Universidade Federal de Goi\'{a}s, Caixa Postal 131, 74001-970, Goi\^{a}nia, Brazil}

\begin{abstract}
We propose a way to encode acceleration directly into quantum fields, establishing a new class of fields.
Accelerated quantum fields, as we have named them, have some very interesting properties. The most important is that they provide a mathematically consistent way to quantize space-time in the same way that energy and momentum are quantized in standard quantum field theories.
\end{abstract}

\maketitle

\section{Introduction}

The description of a classical system from a non-inertial reference frame exhibits fictitious forces. These forces do not stem from any physical interaction, being solely a consequence of the non-inertial feature of the reference frame itself. The notion of fictitious forces also emerge in general theory of relativity. Einstein used accelerated systems and their accompanying fictitious forces, as a guide in order to comprehend the nature of gravity in classical physics. 
Einstein's motivation was the observation that mass never appears in the equations of motion under fictitious forces, a feature which is also encountered with gravity, already known from Galileo's famous experiments.  

Quantum field theory is the best physical model we have in order to describe our world. All the fundamental forces of physics are described within this framework, except one, gravity. Several difficulties arise from the application of the usual prescription of quantum field theory to the quantization of the gravitational field as described by general relativity. This fact led theorists to consider more radical approaches to the problem of quantum gravity. Among them, loop quantum gravity which provides a genuine quantum theory of geometry \cite{Rovelli}.

Einstein concluded that spacetime and the gravitational field are the same physical quantity after recognizing the equivalence between fictitious forces and gravitational forces. Due to the importance of fictitious forces in the geometrization of gravity we believe that fictitious forces could play a key role also in quantization of gravity. A natural question that comes up is the role of these forces in quantum field theory and whether Einstein's idea can also be applied in quantum theory. The investigation of these issues is the main purpose of this paper.
 
As is well known, accelerated systems can be analyzed using special relativity. Since quantum field theories are subject to the rules of special relativity and quantum mechanics, we can ask if there is any quantum field theory capable of embracing acceleration as a fundamental characteristic of the fields. We answer this question in the affirmative by defining the accelerated quantum fields. The structure of accelerated quantum fields permits us to interpret them as fictitious in the sense that they disappear in an inertial reference frame. 

Schrodinger wrote the first relativistic wave equation \cite{Dirac} based on two assumptions: ($i$) the energy-momentum relation for a massive particle is given by $E^2 - p^2 = m^2$ and ($ii$) the particle dynamics should be described by a wave function. Our definition of accelerated fields rests on the kinematical description of the particles and the assumption that their kinematics are described by a different wave function. Our procedure is close to the standard quantization of a free scalar field in flat spacetime and can be shortly described as follows. First, following Schrodinger’s reasoning, we transform the relativistic relation $t^2 - x^2 = -1/ \alpha^2$ into a wave equation in momentum space. Subsequently, after promoting the wave function to a field operator, we solve the field equation and quantize the resulting normal mode structure. 

One of the main results of this paper is the argument that one can define accelerated fields by canonically quantizing the momentum space. Therefore, we need to proceed beyond the standard canonical field quantization in Minkowski spacetime, with the establishment of the field equation and the commutation relations in momentum space and, then, with the construction of the associated framework of observables and quantum states. It is then shown that the emerging notions of particles and vacuum states are different from those of the standard Fock representation. In the theory of accelerated fields, the wave-particle duality appears in the association of the angular frequency and wave vector with space and time, respectively. Here, space and time are quantized, analogously to the way quantities like energy and momentum are quantized in usual quantum field theories.

In a different approach, Sanchez investigated the spacetime structure with quantum theory by promoting the length $x$ and the time $t$ to quantum non-commutative coordinates \cite{Sanchez}. In that work, she found the quantum light cone which is generated by the hyperbolae $x^2 - t^2= \pm [x,t]$ and replaces the classical light cone. Our conclusions regarding accelerated field support such result. Furthermore, keeping in mind that the mathematical language it is used is that of quantum field theory and not that of string theory, our formulation integrates some of the ideas developed in metastring theory \cite{fr-1}.

The idea of quantized spacetime dates back to the early days of quantum theory as a way to eliminate infinities from quantum field theories \cite{Snyder,Carazza} or just to survey the consequences of this assumption \cite{Hill}.
Later it became obvious that quantum spacetime is closely related to proposed theories of quantum gravity.
While it is generally accepted that spacetime is quantized, there is disagreement as to how quantization manifests itself \cite{Smolin}. In this work we advance arguments that inertial acceleration encoded as quantum field provides a mathematical consistent way to quantize spacetime.

By considering a quantum field theory in $1+1$ dimension and adopting the metric signature $(+-)$, our main results are presented in section \ref{main}. We dedicate Sec. \ref{conclusion} for our final discussions, including a possible extension of the theory to $3+1$ dimensions. Furthermore, we shall use units such that $\hbar=c=1$ throughout the paper. Quantities defined in or refereeing to the momentum space will be denoted with a tilde.  

%%%%%%%%%%%%%%%%%%%%%%%%%%%%%%%%%%%%%%%%%%%%
\section{Canonical field quantization in momentum space}
\label{main}

Specific elements in the definition of accelerating fields distinguish them from the massive fields. However, the canonical procedure of quantization we know for scalar fields \cite{Greiner}, appropriately modified, can be applied also to accelerated fields.
 The physical motivation for this is the relativistic invariant relation which describes accelerated particles
\be
x_\mu x^\mu = -\frac{1}{\al^2}.\label{eq.77}
\ee
Inserting the operators for position and time,
\be
t=i\frac{\partial}{\partial E}=i \widetilde{\partial}_E,\qquad x=i\frac{\partial}{\partial p}=i \widetilde{\partial}_p
\ee
into Eq. (\ref{eq.77}) we obtain the differential equation
\be
\left({\widetilde{\partial}_E}^2-{\widetilde{\partial}_p}^2 - \frac{1}{\alpha^2} \right)\wtPsi(p^\mu) = 0,\label{eq.81}\qquad (p^0 = E,\,p^1 = p).
\ee
This is a wave equation with an extra term associated with the proper acceleration. As a wave equation it bears some resemblance to Klein-Gordon equation. The difference arises from the fact that, contrary to Klein-Gordon equation, Eq. (\ref{eq.81}) is associated with the spacelike equation shown in (\ref{eq.77}). We know that each solution of the d'Alembert wave equation $\square \phi(x^\mu) = 0$ is a function in spatial part, $x^i$, and $C^\infty$-dependent on the temporal part, $x^0$, as a parameter. Of course, $\phi(x^\mu)$ can also be regarded as a function in $x^0$ that is $C^\infty$-dependent on $x^i$ as a parameter (see \cite{Bogolubov}). While the set of solutions to the Klein-Gordon equation are parameterized according to the first case, this form of parameterization for the function $\wtPsi(p^\mu)$ leads the obtained frequency to acquire an imaginary part, resulting in an unstable (exponential) growth. Thus, we choose to associate with $\wtPsi(p^\mu)$ the family of functions
\be
\wtPsi_{p}(E) = \left.\wtPsi(p^\mu)\right|_{p^1=p},
\ee
which are $C^\infty$-dependent on $p$ and satisfy the differential equation
\be
\widetilde{\partial_p}^2\wtPsi_{p}(E) = \left({\widetilde{\partial}_E}^2- \frac{1}{\alpha^2}\right)\wtPsi_{p}(E).\label{eq.41}
\ee
The Lagrange density of this equation can be constructed by inverting the Euler-Lagrange equation and is given by
\be
\widetilde{\mathcal{L}} = \frac{1}{2} \left(\left(\widetilde{\partial}_E \wtPsi\right)^2 - \left( \widetilde{\partial}_p \wtPsi\right)^2 +\frac{1}{\al^2}\wtPsi^2\right).\label{eq.76}
\ee
Due to the chosen parameterization of $\wtPsi(p^\mu)$, its uniqueness as solution of (\ref{eq.41}) is satisfied when the initial data $(\wtPsi,\widetilde{\partial_p}\wtPsi)$ are selected on a hypersurface of constant $p$, or a timelike Cauchy surface. So far the field configuration is based only on the relativistic kinematic equation (\ref{eq.77}), the energy-momentum relation has not been encompassed in any way, thus the shape of the fields is supposed not to change under translation in momentum space. Therefore, from the homogeneity of the momentum space, Noether's theorem provides us with a conserved current given by 
\be
\widetilde{\Theta}_{\mu\nu} = \frac{\partial \widetilde{\mathcal{L}}}{\partial (\widetilde{\partial}_\mu \wtPsi)}\widetilde{\partial}_\nu \wtPsi - \eta_{\mu\nu} \widetilde{\mathcal{L}}
\ee
Particularly in our case, for the Lagrangian density (\ref{eq.76}), the quantities that are ``momentum'' independent are
\ba
\widetilde{\Theta}_{pp} &=& \frac{1}{2} \left(\left(\widetilde{\partial}_E \wtPsi\right)^2 + \left( \widetilde{\partial}_p \wtPsi\right)^2 +\frac{1}{\al^2}\wtPsi^2\right) \\
\widetilde{\Theta}_{pE} &=& -\left(\widetilde{\partial}_E \wtPsi\right)\left( \widetilde{\partial}_p \wtPsi\right) \label{Theta_pE}
\ea

Translating this to the Hamiltonian description is a straightforward procedure, we just need to be careful with the peculiarity of our case. The conjugate momentum for $\wtPsi$ is defined as
\be
\wtPi_p(E) =\widetilde{\partial}_p \wtPsi_p(E) ,\label{eq.45}
\ee
thus leading to the Hamiltonian
\be
\widetilde{H} = \int dE \left(\frac{1}{2} (\widetilde{\partial}_E \wtPsi)^2 + \frac{1}{2} \wtPi^2 + \frac{1}{2\al^2}\wtPsi^2 \right),
\ee
in agreement with $\widetilde{\Theta}_{pp}$.

Our next task is to find the spectrum of this Hamiltonian. We do this by finding the solutions of Eq. (\ref{eq.41}). Our guide on this will be the procedure it is followed to solve the Klein-Gordon equation. Using the set of plane waves $e^{i E t}$, we expand $\wtPsi$ as
\be
\wtPsi(p^\mu) = \int d t \, \widetilde{u}_0 \, e^{i E t} \, \wta_t(p),
\ee
with $\widetilde{u}_0$ a normalization factor which will be fixed later. Equation (\ref{eq.41}) then becomes
\be
\widetilde{\partial_p}^2\wta_t(p) = -\left(t^2 + \frac{1}{\alpha^2}\right)\wta_t(p)
\ee
whose general solution can be written as
\be
\wta_t(p) = \wta_t^{(1)} e^{-i p x_t} + \wta_t^{(2)} e^{i p x_t}
\ee
where $\wta_t^{(1)}$ and $\wta_t^{(2)}$ being momentum-independent and the frequency is now defined by the relation
\be
x_t = \sqrt{t^2 + \frac{1}{\alpha^2}}.\label{dis-kin}
\ee
Since we work with real fields, we must have $\wtPsi^\dagger = \wtPsi$. This constraint allows us to write the field operator as 
\be
\wtPsi_p(E) = \int 
\left(
\widetilde{\mathtt{a}}_{t} \widetilde{u}_{t}(E,p) + \widetilde{\mathtt{a}}_{t}^\dagger \widetilde{u}^*_{t}(E,p) 
\right) dt.\label{eq.42}
\ee
where 
\be
\widetilde{u}_{t}(E,p) = \widetilde{u}_0 e^{i (E t- p x_t)}=\widetilde{u}_0 e^{i p^\mu \cdot (x_\mu)'}, \label{eq.591}
\ee 
with $(x^\mu)' = (t,x_t)$, defines the set of orthonormal-mode solutions of (\ref{eq.41}). 

\begin{figure}[htbp]
	\includegraphics[width=0.5\textwidth]{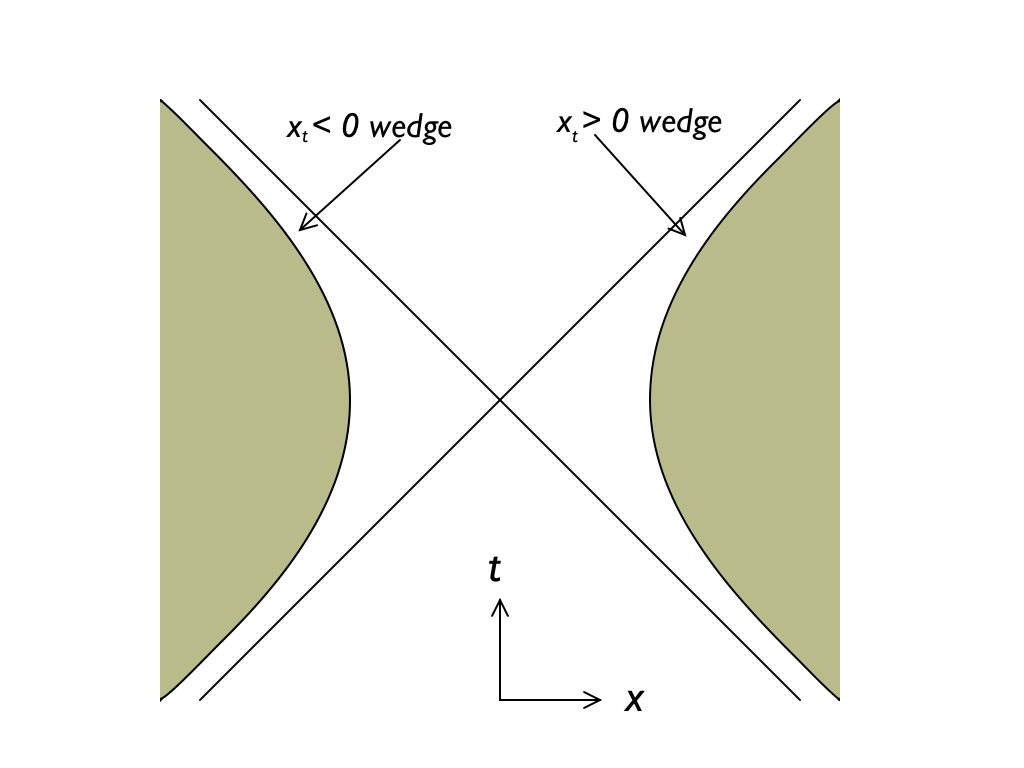}
	\caption{The hyperboloid $x^\mu x_\mu =-1/\al^2$. The two sheets of hyperboloid for a single-particle wavefunction correspond to states of positive and negative position, refering to them more generally as positive- and negative-frequency modes. The Lorentz invariant time integral is over the right wedge of the hyperboloid. 
	}
	\label{fig1}
\end{figure}

For this to make sense, we need to define an inner product on the space of solutions. The appropriate one is expressed as an integral over a constant-momentum hypersurface $\widetilde{\Sigma}$
\be
(\widetilde{\Psi},\widetilde{X}) := i \int_{\widetilde{\Sigma}}    
d\widetilde{\Sigma} \left(\widetilde{\Psi}^* \, \widetilde{\partial}_\mu \widetilde{X} n^\mu
-\widetilde{X} \, \widetilde{\partial}_\mu\widetilde{\Psi}^*n^\mu\right) \label{eq.33}
\ee
with $n^\mu$ a normal vector to the time-like hypersurface $\widetilde{\Sigma}$. 
We can verify that the plane waves (\ref{eq.591}), for 
\be
\widetilde{u}_0 =\frac{1}{\sqrt{4 \pi x_t}},
\ee
with respect to this product, form an orthonormal set
\ba
(\widetilde{u}_{t'},\widetilde{u}_{t}) &=&
\delta(t - t') \\
(\widetilde{u}^*_{t'},\widetilde{u}^*_{t}) &=& 
-\delta(t - t') \\
(\widetilde{u}_{t'},\widetilde{u}^*_{t}) &=&0.
\ea
Note that, like the Klein-Gordon inner product, the inner product (\ref{eq.33}) is not positive definite. From this we can choose the positive-frequency modes as $\widetilde{u}_t$'s and, consequently, the negative-frequency modes as $\widetilde{u}^*_t$'s. It is important to emphasize that in this construction, unlike the Klein-Gordon case, the division of the modes into positive- and negative-frequency occurs with the spatial dependence of the type $e^{- i p x_t}$ and $e^{i p x_t}$, respectively (see Fig.\ref{fig1}).

%%%%%%%%%%%%%%%%%%%%%%%%%%%%
%%%%%%%%%%%%%%%%%%%%%%%%%%%%

If we define the annihilation and creation operators associated with $\widetilde{u}_{t}$ and $\widetilde{u}_{t}^*$ by the relations
\be
\widetilde{\mathtt{a}}_{t} = (\widetilde{u}_{t},\wtPsi) \hspace{0.5cm} \mbox{and} \hspace{0.5cm} \widetilde{\mathtt{a}}_{t}^\dagger = -(\widetilde{u}^*_{t},\wtPsi) \label{eq.57},
\ee
we can write $\wtPsi$ and $\wtPi$ as a linear combination of an infinite number of creation and annihilation operators $\widetilde{\mathtt{a}}_{t}^\dagger$ and $\widetilde{\mathtt{a}}_{t}$, indexed by the time $t$ as
\be
\wtPsi_p(E) = \int \frac{dt}{\sqrt{2 x_t (2\pi)}} \,  \left(\widetilde{\mathtt{a}}_{t} e^{i p^\mu \cdot (x_\mu)'} + \widetilde{\mathtt{a}}_{t}^\dagger e^{-i p^\mu \cdot (x_\mu)'}\right)\label{eq.59}
\ee
and
\be
\wtPi_p(E) = -i \int \frac{dt}{\sqrt{2\pi}} \sqrt{\frac{x_t}{2}} \left(\widetilde{\mathtt{a}}_{t} e^{ i p^\mu \cdot (x_\mu)'} - \widetilde{\mathtt{a}}_{t}^\dagger e^{-i p^\mu \cdot (x_\mu)'}\right) . \label{eq.59A}
\ee

The operators $\widetilde{\mathtt{a}}_{t}$ and $\widetilde{\mathtt{a}}_{t}^\dagger$ should fulfill the typical algebra for creation and annihilation operators, i.e. 
\be
\left[\widetilde{\mathtt{a}}_{t},\widetilde{\mathtt{a}}_{t'}\right] = \left[\widetilde{\mathtt{a}}_{t}^\dagger,\widetilde{\mathtt{a}}_{t'}^\dagger\right] = 0,\quad \left[\widetilde{\mathtt{a}}_{t},\widetilde{\mathtt{a}}_{t'}^\dagger\right] = \delta(t-t') \label{eq.61}
\ee
Due to Eqs. (\ref{eq.59}) and (\ref{eq.59A}), the commutation relations (\ref{eq.61}) in coordinate space are equivalent to the equal-momentum canonical commutation relations
\ba
\left[\wtPsi_p(E),\wtPsi_{p}(E')\right] &=& \left[\wtPi_p(E),\wtPi_{p}(E')\right] =0 \\
\left[\wtPsi_p(E),\wtPi_{p}(E')\right] &=& i \delta(E - E'),
\ea
in momentum space. This is the first important result of our formulation. It might seem disturbing at first sight, but it is consistent with the nature of the wave equation we employed to quantize the field in the momentum space.
 
After we have defined the creation and annihilation operators, we can then express the Hamiltonian in terms of those operators
\be
\widetilde{H} = \int \frac{dt}{\sqrt{2 x_t (2\pi)}} \,  x_t\left(\widetilde{\mathtt{a}}_{t}^\dagger\,\, \widetilde{\mathtt{a}}_{t} +\frac{1}{2}\left[\widetilde{\mathtt{a}}_{t},\widetilde{\mathtt{a}}_{t}^\dagger\right]\right) \label{Ham-pos}
\ee
The second term is proportional to $\delta(0)$, an infinite c-number. It is the sum over all modes of the zero-point ``positions'' $x_t/2$. We cannot avoid the existence of this term, since our treatment resembles that of harmonic oscillator and the infinite c-number term is the field analogue of the harmonic oscillator zero-point energy. Considerations coming from the standard field theory suggests discarding this term.

Using Eq. (\ref{Ham-pos}) for the Hamiltonian, it is straightforward to evaluate the commutators
\be
\left[\widetilde{H},\widetilde{\mathtt{a}}_{t}^\dagger\right] = x_t \widetilde{\mathtt{a}}_{t}^\dagger,\qquad \left[\widetilde{H},\widetilde{\mathtt{a}}_{t}\right] = -x_t \widetilde{\mathtt{a}}_{t}.
\ee
We can then write down the spectrum of the theory. There will be a single vacuum state $\ket{0_{\widetilde{\mathtt{a}}}}$, characterized by the fact that it is annihilated by all $\widetilde{\mathtt{a}}_t$,
\be
\widetilde{\mathtt{a}}_t \ket{0_{\widetilde{\mathtt{a}}}} =0,\qquad  \forall t.
\ee 
All other eigenstates can be built by letting $\widetilde{\mathtt{a}}_t^\dagger$ acting on the vacuum. Let
\be
\ket{t} = \widetilde{\mathtt{a}}_t^\dagger\ket{0_{\widetilde{\mathtt{a}}}}
\ee

Having found the spectrum of the Hamiltonian, let us try to interpret its eigenstates. By similar logic applied to the Hamiltonian $\widetilde{H}$, we can construct an operator corresponding to the one showed in Eq. (\ref{Theta_pE})
\be
\widetilde{T} = - \int dE\,\, \widetilde{\mathtt{\Pi}} \widetilde{\partial}_E \widetilde{\mathtt{\Psi}} = \int \frac{dt}{\sqrt{2 x_t (2\pi)}} \,  t\,\,\widetilde{\mathtt{a}}_{t}^\dagger\,\, \widetilde{\mathtt{a}}_{t}  \label{T-time}
\ee 
So the operator $\widetilde{\mathtt{a}}_{t}^\dagger$ creates excitations at time $t$ and position $x_t = \sqrt{t^2 + 1/\alpha^2}$. These excitations are discrete entities that have the proper relativistic kinematics relation which describes uniform accelerated particles. We claim that we can call these excitations accelerating particles and the associated fields as accelerated fields. By a particle here we mean something that is localized in space, since $\widetilde{\mathtt{a}}_{t}^\dagger$ creates space eigenstates. But this operator says nothing regarding the momentum of the particle. This information can be derived considering the interpretation of the state $\widetilde{\mathtt{\Psi}}(p^\mu)\ket{0_{\widetilde{\mathtt{a}}}}$. From the expansion (\ref{eq.59}) we see that
\be
 \wtPsi(p^\mu) \ket{0_{\widetilde{\mathtt{a}}}} = \int dt \,\,
 \widetilde{\mathtt{u}}_t^*(p^\mu)  \,\, \widetilde{\mathtt{a}}_{t}^\dagger \ket{0_{\widetilde{\mathtt{a}}}} \label{eq.69}
\ee
is a linear superposition of single particle states that have well-defined time-position, hence we will claim that the operator $\wtPsi(p^\mu)$, acting on the vacuum, creates a particle with momentum $p$ and energy $E$. Note that, since we considered an hermitian field operator, $\wtPsi^\dagger(p^\mu)$ creates particles with the same momentum and energy as $\wtPsi(p^\mu)$ does.
Therefore, we quantized the accelerated field in the Schrödinger picture and interpreted the resulting theory in terms of accelerated relativistic particles. In the Heisenberg picture, the operator $\widetilde{X}^\mu = (\widetilde{T},\widetilde{H})$ allows us to relate $\widetilde{\mathtt{\Psi}}(p^\mu)$ to $\widetilde{\mathtt{\Psi}}(0)$ by means of the relation
\be
\widetilde{\mathtt{\Psi}}(p^\mu) = e^{-i \widetilde{X}^\mu p_\mu}\, 
\widetilde{\mathtt{\Psi}}(0) \, e^{i \widetilde{X}^\mu p_\mu}.
\ee

For one particle state of well-defined time, $t$ is the eigenvalue of $\widetilde{T}$. Equation (\ref{eq.59}) makes explicit the dual particle and wave interpretations of the quantum field $\widetilde{\mathtt{\Psi}}$. On the one hand, $\widetilde{\mathtt{\Psi}}$ is written as a Hilbert space operator, which creates particles that are the quanta of the field. On the other hand, $\widetilde{\mathtt{\Psi}}$ is written as a linear combination of the solutions $e^{\pm ip^\mu\,x_\mu}$ of the wave equation (\ref{eq.41}). If these were single-particle wavefunctions, they would correspond to states of positive and negative position; We have chosen to refer to them more generally as positive- and negative-frequency modes, keeping in mind the difference with the corresponding modes coming from Klein-Gordon equation. The connection between the particle creation operators and the waveforms has already been mentioned above. A positive-frequency solution of the field equation has as its coefficients the operator that destroys a particle in that particular mode. While a negative-frequency solution, being the Hermitian conjugate of a positive-frequency solution, has as its coefficient the operator that creates a particle in that positive-position single-particle wavefunction.

Let us consider  a real scalar field described by the Klein-Gordon equation $(\square + m^2)\psi(x^\mu) = 0$. To illustrate the difference between the Klein-Gordon field and the field theory we developed from the kinematic wave equation (\ref{eq.81}) let us compute the quantities $\bra{0}\psi(x^\mu)\ket{p}$ and $\bra{0_{\widetilde{\mathtt{a}}}}\widetilde{\mathtt{\Psi}}(p^\mu)\ket{t}$, which can be interpret as the position representation of the single-particle wavefunction of the state $\ket{p}$ and the momentum representation of the state $\ket{t}$, respectively. We find
\ba
\bra{0}\psi(x^\mu)\ket{p} \equiv \braket{x^\mu}{p} &=& \frac{1}{\sqrt{4 \pi E_p}} \left.e^{-i \,p^\mu \cdot x_\mu}\right|_{p^0 = E_p} \\
\bra{0}\widetilde{\mathtt{\Psi}}(p^\mu)\ket{t} \equiv \braket{p^\mu}{t} &=& \frac{1}{\sqrt{4 \pi x_t}} \left.e^{i \,p^\mu \cdot x_\mu}\right|_{x^1 = x_t}
\ea
In the first case we recognize the relativistic dispersion relation for a particle with mass $m$, while in the second one we identify the relativistic kinematic relation for uniformly accelerated particle.

After the introduction of Heisenberg picture, we can proceed to the calculation of the amplitude for a particle to propagate from $k^\mu$ to $p^\mu$, which in our present formalism is $\bra{0_{\widetilde{\mathtt{a}}}}\widetilde{\mathtt{\Psi}}(p^\mu)\widetilde{\mathtt{\Psi}}(k^\mu)\ket{0_{\widetilde{\mathtt{a}}}}\equiv \left\langle \widetilde{\mathtt{\Psi}}(p^\mu)\widetilde{\mathtt{\Psi}}(k^\mu) \right\rangle_0$. We will call this quantity $\widetilde{D}(p^\mu - k^\mu)$ and we easily check that it is given by the integral
\be
\widetilde{D}(p^\mu - k^\mu) \equiv \left\langle \widetilde{\mathtt{\Psi}}(p^\mu)\widetilde{\mathtt{\Psi}}(k^\mu) \right\rangle_0
= \int \frac{dt}{\sqrt{2\pi}2 x_t}\,e^{i\, \left(x_\mu \right)'(p^\mu - k^\mu)} .
\ee  
Let us evaluate this integral for some particular values of $p^\mu - k^\mu$.

First consider the case where the difference is: $p^0-k^0=0$, $p^1-k^1=\pm \, q$. Then we have
\ba
\widetilde{D}_{\pm}(p^\mu - k^\mu) &=& \frac{1}{\sqrt{2\pi}}\int \frac{dt}{2 x_t}\,e^{\mp i\, x_t \,q} \nonumber \\
&=& \frac{1}{\sqrt{8\pi}}\int \frac{dx_t}{ \sqrt{x_t^2 - 1/\alpha^2}}\,e^{\mp i\, x_t\, q} \nonumber \\
&=& \sqrt{\frac{1}{2\pi}} \mathcal{K}_0\left(\mp\frac{i\,q}{\alpha}\right) \nonumber \\
&\stackrel{q \rightarrow \infty}{\sim}& e^{\mp\frac{i\,q}{\alpha}}
\ea
where $\mathcal{K}_\nu(x)$ is the modified Bessel function for $\nu =0$. 

Next consider the case where $p^0-k^0=\pm\mathcal{E}$, $p^1-k^1=0$. The amplitude is
\ba
\widetilde{D}_\pm (p^\mu - k^\mu) &=& \frac{1}{\sqrt{2\pi}}\int \frac{dt}{2 \sqrt{t^2 + 1/\alpha^2}}\,e^{\pm i\, t \, \mathcal{E}} \nonumber \\
&=& \sqrt{\frac{1}{2\pi}} \mathcal{K}_0\left(\frac{\mathcal{E}}{\alpha}\right) \nonumber \\
&\stackrel{\mathcal{E} \rightarrow \infty}{\sim}& e^{\pm\frac{\mathcal{E}}{\alpha}}
\ea
We find that the propagation amplitudes are not zero for particles propagating over all intervals, timelike and spacelike. 

\begin{figure}[htbp]
	\includegraphics[width=0.5\textwidth]{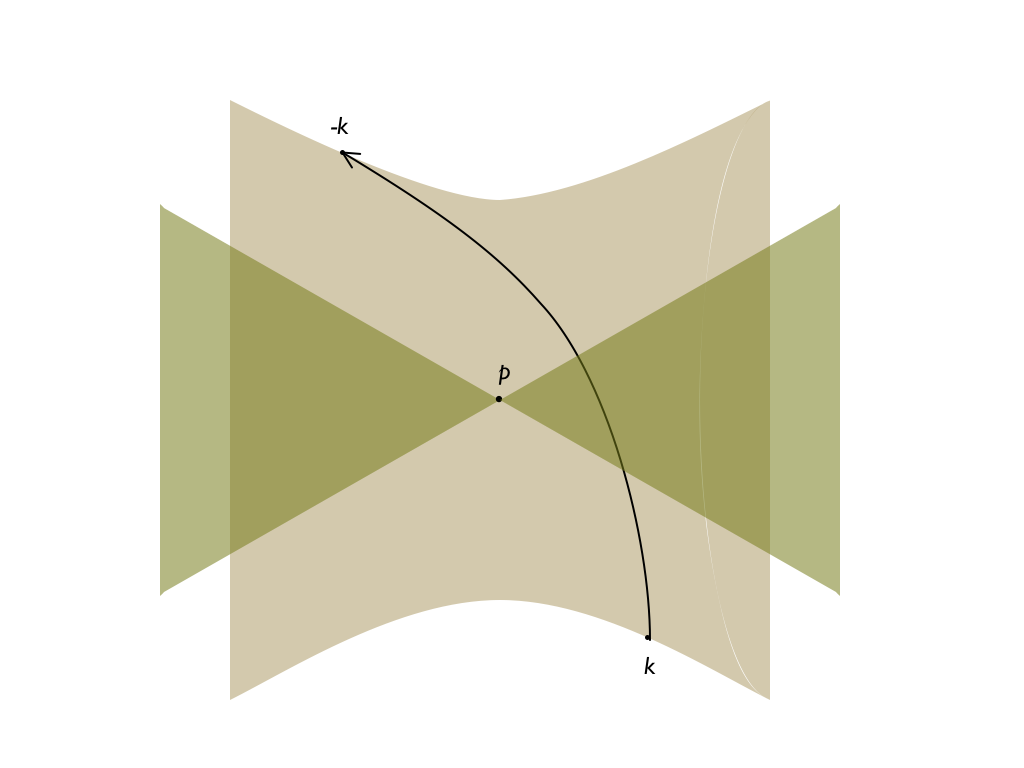}
	\caption{When $p^\mu -k^\mu$ is timelike, a continues Lorentz transformation can take $p^\mu -k^\mu$ to $-(p^\mu -k^\mu)$. }
	\label{fig2}
\end{figure}

A measurement performed at one point can affect a measurement at another point if the associated commutator does not vanish. Thus, a theory, in order to be causal, should require all the operators defined outside the light-cone to commute. Let us compute the commutator $[\widetilde{\mathtt{\Psi}}(p^\mu),\widetilde{\mathtt{\Psi}}(k^\mu)]$ whose general form is given by 
\be
[\widetilde{\mathtt{\Psi}}(p^\mu),\widetilde{\mathtt{\Psi}}(k^\mu)] 
= \widetilde{D}(p^\mu - k^\mu) - \widetilde{D}(k^\mu - p^\mu). \label{eq.comm}
\ee
As shown in Fig. \ref{fig2}, when $(p^\mu - k^\mu)^2>0$, being outside the light-cone, an appropriate continuous Lorentz transformation can take us from $p^\mu - k^\mu$ to $-(p^\mu - k^\mu)$. If we perform this transformation on the second term in (\ref{eq.comm}), the two terms inside parenthesis are therefore equal and cancel. In case $(p^\mu - k^\mu)^2<0$, there is no continuous Lorentz transformation that can take $p^\mu - k^\mu$ to $-(p^\mu - k^\mu)$, thus the amplitude is nonzero. We then conclude that no measurement in our theory of accelerated fields can affect another measurement outside the light-cone. Hence, causality is maintained. We intentionally didn't discuss causality by recruiting spacelike/timelike intervals. Contrary to Klein-Gordon theory, our field commutator vanishes for timelike separations while survives for spacelike separations. This outcome is consistent with our choice the role of ``time'' to be played by the spacelike coordinate $p$, resulting in a light-cone with different direction (see Fig. \ref{fig1}) and interchanged notions of timelike and spacelike intervals.

It is crucial, in order to satisfy the causality requirement, that we quantize the accelerated field using commutators. Had we used anticommutators instead of commutators, we would get a nonvanishing result for timelike intervals. Thus, this is an indication that the ``spin-statistics theorem'' also holds for relativistic particles. Another crucial element in our construction is essential for causality: The negative-space eigenfunctions in the field operators. Again, without negative-space contributions Eq.(\ref{eq.comm}) does not vanish for spacelike separations. 

%%%%%%%%%%%%%%%%%%%%%%%%%%%%%%%%%%%%%%

\section{Discussion}
\label{conclusion}

Although we have considered field theories in $1+1$ dimensions, our results can be extended to physical dimensions. In terms of a global inertial coordinate system $t,x,y,z$, let us consider the killing field which generates a boost about the origin in the $x$ direction. In this case, the hyperbolic surface $t^2-x^2=-1/\alpha^2$ is invariant under translation in $y$ and $z$ direction. Given the fact that the definition of accelerated fields is based on this hyperbolic cylinder surface, in our theory, accelerated fields become invariant under translations in $y$ and $z$ direction. This clearly reproduces all the results presented above. In the general case, we will have three dimensional surface that will be projected into the planes ($t,x$), ($t,y$) and ($t,z$), thus defining three independent accelerated fields. This reflects the fact that, in our theory, geometry is translated into fields.

Even though gravity, as described in general relativity, appears not to be compatible with the quantum field theories, this is not the case for the accelerated fields. In this work we propose a new quantum field theory, that encodes acceleration directly into the fields and provides a mathematical consistent way to locally quantize spacetime. This is a strong evidence that not only gravity but also acceleration can reshape the notions of space and time.

The problem of finding a quantum theory of gravity is the main problem in modern theoretical physics. Today, the two main approaches are string theory and loop quantum gravity. These theories address the problem of quantum gravity in totally different ways. String theory is based on the idea of a fundamental description of gravity in terms of physical excitations of a string that lives in a background metric space. Loop quantum gravity, on the other hand, in an attempt to comprehend what quantum spacetime is at the fundamental level, was formulated without a background spacetime. 

The theory of accelerated fields shares elements from both of these theories. Even though the description of acceleration takes place in terms of excitations over a background metric space, the fact that this space is the momentum space allows the quantization of spacetime, the background of all the others fields. From this view one can define accelerated Klein-Gordon fields establishing a notion of scalar fields being located with respect to the accelerated field.

The theory of accelerated fields, as presented here, is an effort to quantize spacetime, while it is kept separate from the matter fields. In a subsequent study, we intend to investigate the effects of accelerated fields on the Klein-Gordon massive field, by defining accelerated Klein-Gordon massive fields and comparing them with the inertial ones.

Several other directions follow from this work. For instance, the question of understanding entanglement in non-inertial frames has been central to the development of the emerging field of relativistic quantum information \cite{P-T,Fuentes,Celeri}. The approach presented in this work offers a new ground to address this fundamental question from a completely different point of view. 

\begin{acknowledgments}
We would like to thanks George E. A. Matsas, Daniel Terno and Daniel A. T. Vanzella for discussions. We acknowledge financial support from the Brazilian funding agencies CNPq (Grants No. 401230/2014-7, 305086/2013-8 and 445516/2014-) and CAPES (Grant No. 6531/2014-08), the Brazilian National Institute of Science and Technology of Quantum Information (INCT/IQ).
\end{acknowledgments}

\end{document}